\documentclass[conference, 10p]{IEEEtran}
\usepackage[utf8]{inputenc}
\usepackage[T1]{fontenc}
\usepackage{graphicx}
\usepackage{longtable}
\usepackage{wrapfig}
\usepackage{rotating}
\usepackage[normalem]{ulem}
\usepackage{amsmath}
\usepackage{amssymb}
\usepackage{capt-of}
\usepackage{hyperref}
\usepackage{graphicx}
\usepackage{url}
\usepackage{multirow}
\usepackage{fancyhdr}
\usepackage{subcaption}
\usepackage{xcolor}
\usepackage{pbox}
\usepackage[boxruled]{algorithm2e}
\hypersetup{draft}
\author{Hermann Kopetz, TU Wien, Austria \\ Wilfried Steiner, TTTech, Austria}
\date{\today}
\title{Temporal Consistency of Data and Information in Cyber-Physical Systems}
\hypersetup{
 pdfauthor={Hermann Kopetz, TU Wien, Austria \\ Wilfried Steiner, TTTech, Austria},
 pdftitle={Temporal Consistency of Data and Information in Cyber-Physical Systems},
 pdfkeywords={},
 pdfsubject={},
 pdfcreator={Emacs 28.1 (Org mode 9.5.2)}, 
 pdflang={English}}
\begin{document}

\maketitle
\begin{abstract}
In a large cyber-physical system, a temporal inconsistency of an output value can arise if there is a non-negligible delay between the instant when a sensor value is acquired from the environment and the instant when a setpoint, based on this sensor value, is used in the environment. Such a temporal inconsistency can be the cause of a critical malfunction of the cyber-physical system. This paper presents a solution of this temporal consistency problem that can best be implemented in a time-triggered architecture (TTA). In a TTA, the instants of sensor value acquisition, setpoint calculation, and actuation on the environment are statically configured, and the cyber-physical system implements software and hardware mechanisms to execute the respective actions tightly at these configured instants. 
\end{abstract}

\section{Introduction}
\label{sec:orgab16478}
\label{org76ed55f}

A cyber-physical system (CPS) consists of two interacting subsystems: a physical system (PS), e.g., a mechanical machine that operates in the real world, and a controlling computer system, the cyber-system (CS) that operates in cyber-space and controls the operation of the PS. Examples of a CPS are an industrial production system, an autonomous vehicle, a smart grid, or a smart building.  A temporal inconsistency can arise in a CPS when there is a non-negligible delay between the instant when a sensor value is acquired from the environment (the PS) and the instant when a setpoint, based on this sensor value, is used in the environment of the CS, (the PS). A temporally inconsistent setpoint produced by the CS can lead to a serious, possibly catastrophic, failure in the PS.

The objective of this paper is to provide a solution to the problem of temporal inconsistency of an output information item, e.g., an inconsistent setpoint, produced by the CS.

The problem of temporal consistency of real-time information has already been addressed more than 25 years ago, particularly in the context of databases, by Xiong et al.  \cite{xiong1996maintaining} and later by Kuo and Mok  \cite{kuo2000real} who states on p.1241: \emph{"External consistency requirements keep real-time databases up-to-date; temporal consistency requirements ensure that multiple data objects read by a transaction are compatible in the currency of their data."}  Our notion of temporal consistency covers both aspects: the up-to-date aspect and the compatibility aspect.  In a recent extensive literature survey on Data Quality Challenges in Large Scale Cyber-Physical Systems, Alwan et al. \cite{alwan2022data} assert that further research is required to address the challenge of timeliness and temporal consistency in cyber-physical systems.

The foundation of this paper is the \emph{Data, Information and Time} (DIT) model \cite{kopetz2022data} that makes a clear distinction between a \emph{data item} and an \emph{information item}. In the real world, a data item is a symbol, where the signifier of the symbol is an acoustic or visual pattern (e.g., a word), and the signified is the \emph{meaning} of this pattern in the mind of a human. An information item is the \emph{sense} of a natural language proposition that consists of multiple linked data items and a timing parameter that tells when the information item is valid.

This paper is organized into nine Sections. The following Section \ref{org18ce284} introduces the system concepts and explains some of the technical terms, particularly related to time, that are used in the rest of the paper. The fundamental distinctions between the properties of the PS, which operates in the time-sensitive real world, and the properties of the CS, which operates in the quasi-timeless domain of mathematics and logic, are elaborated. Section \ref{orgaf2545e} introduces the notions of a data item and an information item. In Section \ref{orgcbb5c25}, a definition of the concept of temporal consistency of an information item is given. An example from the automotive domain shows that an inconsistent information item can have catastrophic consequences. Section \ref{orgd66a559} presents a solution to this problem that is supported by a time-triggered system architecture. Section \ref{orgfe587a8} discusses the co-simulation of a CPS. Section \ref{org5372be1} presents six design principles that should be considered in the design of a CPS to avoid a temporal inconsistency of the output data. Section \ref{org0c165d0} discusses related work and the paper closes with a short conclusion in Section \ref{org3e261b7}.

\section{System Concepts}
\label{sec:org7a6d67f}
\label{org18ce284}
By the term \emph{system}, we mean a \emph{whole} that serves a given purpose and has a \emph{real or virtual skin} that separates the whole from its environment.  In the skin of a system, there are input and output interfaces that provide the connections between a system and its environment --- in the case of the CPS, the \emph{sensors} and \emph{actuators} establish the links between the CS and the PS.  A sensor generates an observation by observing some part of the state of its environment, the PS, at a given \emph{interaction-instant}. An actuator forces the result of a computation executed in the CS on its environment (the PS) at an assumed \emph{interaction-instant}. There are two very different models of time in the PS and the CS.

\subsection{PS model of time}
\label{sec:orgcb348c5}

The applicable model of time, the \emph{physical real-time}, in the PS, is based on Newtonian physics, which considers time to be an absolute quantity that progresses continuously and autonomously from the past to the future. We represent the progress of time metaphorically by a directed timeline and call a cut of the timeline an \emph{instant}. A digital clock partitions the timeline by its ticks (instants) into a sequence of uniformly numbered sections, the \emph{granules} of the clock, where each granule is delimited by two ticks of the clock, the start tick of the granule and the start tick of the following granule. A \emph{timestamp} of an instant is the number of the start tick that occurred immediately before the instant. The starting moment for the numbering of the granules, measured in seconds, is the epoch (in the \emph{Coordinated Universal Time Standard (UTC)} the epoch is midnight of Jan. 1, 1970). An interval is a section of the timeline between the start instant of the interval and the termination instant of the interval. A duration is the length of an interval expressed in the number of contained clock granules. We call the interval that is of relevance for our considerations the \emph{Interval of Discourse (IoD)}. If a property of an entity of the PS does not change during the IoD, we call the property \emph{static}, otherwise \emph{dynamic}. In the real world, most properties are dynamic.  A happening that occurs during an interval that is shorter than the granule of the clock is called an \emph{event}. The timestamp of the event is the start tick of the clock granule that preceded the event.  (For a more detailed discussion on time measurement with digital clocks, see \cite{kopetz2022real}).

\subsection{CS model of time}
\label{sec:org108558e}

The model of time in the CS is not continuous but discrete. In a discrete model, real-time is partitioned into a sequence of intervals --- we call such an interval a \emph{computation-interval} --- where a computation-interval starts at an interaction instant, i.e., an instant where the PS and the CS exchange information, and terminates at the following interaction instant. A  change of real-time takes place in the CS only at an interaction instant.  During a computation interval, the real-time in the CS does not advance. The reason for this discrete model of time lies in the nature of mathematical and logic operations that are executed by the computer during a quasi-timeless \emph{computation-interval}.  The variables and operators of mathematics and logic are timeless universals \cite[Chapter~9]{russell2001problems}. The result of the execution of the algorithm 3+4 = 7 does not depend on any specific duration, any specific time of execution, or the place of execution. The only timing issue in the CS system concerns the duration of the execution time of an algorithm, but this timing issue does not affect the formal correctness of an algorithm. If a result (a new setpoint) must be provided to the PS at a particular interaction instant, the chosen computer must be performant enough to terminate the execution of the algorithms before this particular interaction instant\footnote{The computer may also be a distributed computer system. The IEEE Technical Community on Real-Time Systems researches methods to calculate worst-case time bounds for CS computations. The application of said methods determines if a computer's performance is adequate.}. A discrete CS model of time, in which real-time progresses stepwise, is also used in discrete event simulations by introducing the notion of an event calendar \cite{dutertre2004modeling}.

In order to simplify the handling of the progress of real-time in the CS, a (temporal) \emph{frame structure} is often introduced in the CS. A frame structure partitions real-time into a regular sequence of uniform intervals, called the \emph{frames}, where a frame starts at the instant Start-Of-Frame (SOF) and terminates at the end of a frame, which is the SOF of the following frame. In the CS, a change in real-time takes place only at the instant start of a frame (SOF). During a frame, the local time does not advance until the SOF of the next frame occurs (which indicates the termination of the current frame). \textbf{Interactions between the PS and CS are only allowed to occur at SOFs.} Typically, a computation in the CS starts by acquiring a data item from the PS at an SOF of a frame, computes the result during the frame, and provides an output (setpoint) to the PS at the end of the frame (which is the SOF of the following frame).  The frame model simplifies the handling of physical time in the CS but has its limitations when the temporal latency of a sensor or the temporal latency of an actuator must be considered, as shown in Section \ref{orgd66a559} of this paper. 

\subsection{Data and Information in PS vs. CS}
\label{sec:orgc97f4c3}

Another fundamental difference between the PS and the CS concerns the denotation of the terms \emph{data} and \emph{information}. In the real world (the PS), a \emph{data item} is a symbol that refers to the \emph{meaning} of the named concept in the mind of an involved human. An information item is the \emph{sense} (the idea or intention) communicated by a natural language proposition to a human receiver at the given instant of utterance. Such an \emph{information item} (which we call ITOM) is formed by an atomic complex that relates multiple data items and is valid at a specified instant or during a specified interval. In the CS, a data item is a timeless value (a bit pattern) that denotes a universal --- e.g., a bare number --- in the domain of mathematics. Multiple values can be connected by timeless mathematical operators to express an algorithm that can be executed on a computer. If a variable of the CS has a name that has significance in the PS --- by relating the value of the variable to the named concept that has meaning in the real world of the PS --- we say that the data item (the value) is \emph{grounded} \cite{pavlick2023symbols}.  A grounded data item, i.e., a computer variable with a name that explains the meaning of the variable in the PS, connects the value stored in a memory location of the CS to a concept in the mind of a human in the PS. It is thus an information item (see below in Section \ref{orgaf2545e} and \ref{orgcbb5c25}).

Natural language propositions, formed by the concatenation of natural language words and augmented by formal mathematical models, are used to express the \emph{specification} of a CPS.  The meaning of a natural language word used in human communication is, on the one hand, anchored in the subjective \emph{conceptual landscape} of the speaker and, on the other hand, interpreted by the listener using his/her subjective conceptual landscape.  This \emph{subjective conceptual landscape} of a human is the result of a lifelong learning process. It contains subjective concepts (units of thought) and many personal mental models of the perceived and learned regularities (laws) that are assumed to govern the processes in the real world. These \emph{subjective conceptual landscapes} of a speaker and a listener are not ident. The resulting ambiguity of natural language words and the ongoing changes, in reality, can make the specification of a real-world system imprecise. In contrast, the values and algorithms contained in the specification of the CS are well-defined in the formal world of mathematics, and the execution of an algorithm is precise under the assumption that the computer hardware works correctly.

\section{Data and Information}
\label{sec:org61a185d}
\label{orgaf2545e}

In colloquial --- and even in scientific --- language, the terms data and information are not precisely defined. Sometimes they are used as if they were synonyms and some other times information is considered to be processed data, without explaining what is the precise difference between raw data and processed data. In this section we present the view that \emph{data} and \emph{information} are different concepts and try to show the remarkable difference. 

\subsection{Data Item in the Real World (the PS)}
\label{sec:orgd2428ee}
\textbf{In the real world, a \emph{data item} is a \emph{symbol} where the \emph{signifier} is a physical pattern, e.g., a \emph{sequence of letters}, an \emph{acoustic utterance}, or a \emph{gesture}, and the \emph{signified} is the meaning that is assigned to this pattern by a human.}

We distinguish between three different signifiers in a real-world communication act: 1) Proper name, 2) Literal, and 3) Natural Language Word.

\subsubsection{Proper Name as Signifier}
\label{sec:orge9f0a54}
A \emph{proper name} acquires its meaning by designating a particular entity, e.g., a person, a thing, a construct, or a memory location in a computer, as its \emph{referent} (i.e., the entity that it refers to) \cite{kripke1980naming}.

\medskip
\textbf{Example:} In a computer program, the name of the variable xyz is a signifier that denotes a memory location in the computer as its signified (as its referent). \P
\medskip

Since the \emph{signifier} xyz has no meaning outside its limited context of the computer program, it is called a \emph{bare name}.

There is a possible second function of a name, which is to provide an explanation of the named (the referent) by linking the referent to a real-world concept \cite[p.26]{quine2011}.

\medskip
\textbf{Example:} The blue van from the garage. The first function of this phrase is to designate a particular car, while the second function is to explain the type and place of the car. \P
\medskip

A signifier that is used as a name and has a meaning in the real world is called a \emph{grounded name}. \emph{Grounding a name} refers to the process of establishing a link between a \emph{bare name} in the CS and an entity in the PS \cite{pavlick2023symbols}.

It is important to separate these two functions of a name from each other and to be aware when a name does not provide the second function of explaining. Bare names are timeless \emph{universals} that are widely deployed in formal systems. While the meaning of a bare name in a formal system is static, the meaning of a grounded name can change when the referred real-world entity has changing properties.

\subsubsection{Literal as Signifier}
\label{sec:org5806fde}
\begin{figure}[htbp]
\centering
\includegraphics[width=8cm]{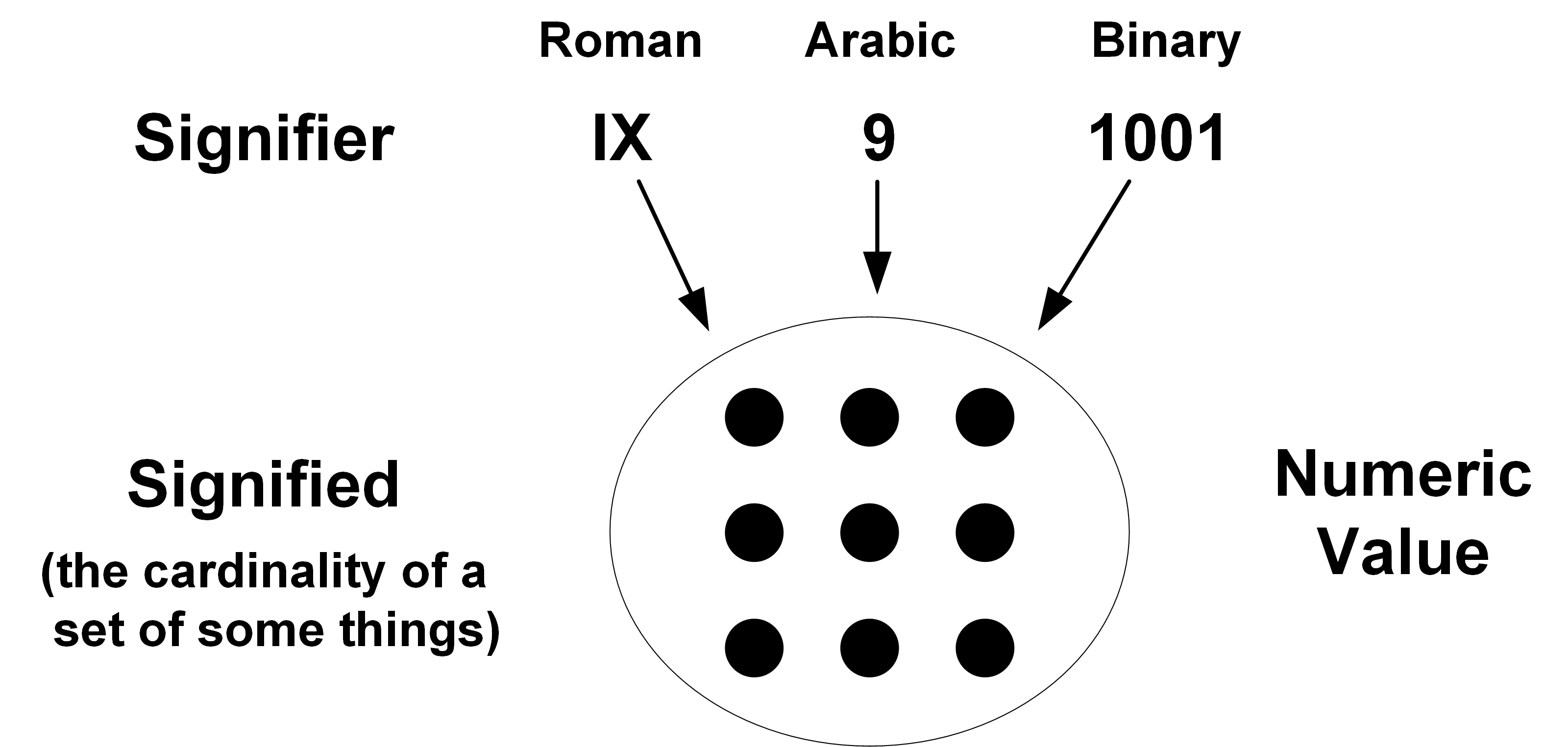}
\caption{\label{fig:org10d8afe}Numeric value of a literal}
\end{figure}

A literal is a pattern where, in a given context, the meaning of the pattern, the \emph{signified}, can be derived from the gestalt of the pattern, the \emph{signifier}.  Depending on the cultural context, different signifiers can have the same meaning (see Fig. \ref{fig:org10d8afe}). The meaning of literals is always timeless and therefore \emph{static}. Literals are widely used to denote \emph{constant numeric values}.

\medskip
\textbf{Example:} If the agreed context is \emph{binary numbers}, the signifier \emph{1001} means the numeric value \emph{9}, expressed in the \emph{western Arabic numeral} system. \P
\medskip

To get an understanding of the significance of a numeric value of the CS in the PS, we need an \emph{explanation} that grounds the numeric value in the real world by providing a \emph{measurement unit} and a meaningful property name of an entity in the PS.

In the physical world, a \emph{measurement} consists of a \emph{numeric value} complemented by the applicable \emph{measurement unit}, called the \emph{dimension} of the value.  The dimension explains the value by denoting an agreed measurement system, e.g., the international SI system.  A bare numeric value, without an identification of the dimension, is of no use in the PS. Care must be taken when performing arithmetic operations on the numerical values of measurements. The \emph{addition} and \emph{subtraction} of numerical values are only permitted if the measurements have the same dimension, which will also be the dimension of the result. The \emph{multiplication} or \emph{division} of numerical values that have different dimensions result in a new value with a \textbf{new} dimension.

\subsubsection{Natural language word as Signifier}
\label{sec:org822bee8}
The meaning of a \emph{natural language word} is the \emph{acquired concept, i.e., the unit of thought}, that is associated with the word in the \emph{conceptual landscape}, the mind, of the language user. The issue of concept formation, i.e., how a word acquires its meaning, is discussed in \cite[pp.17-20]{kopetz2022data}.

\bigbreak

Let us look at the following example where two static data items are linked by the predicate \emph{is}:

\medskip
\textbf{Example:} The \emph{engine} is \emph{overheated}. \P
\medskip

Although the meanings of the terms \emph{engine} and \emph{overheated} are static, the proposition formed by these static terms is dynamic, i.e. it is valid only at given a point in time or during a given interval. The word \emph{is} brings the temporal aspect --- the present --- into play.

\subsection{Data Item in Cyberspace (the CS)}
\label{sec:orgf281d94}
\textbf{In cyberspace, a \emph{data item} is a symbol where the \emph{signifier} is a bit pattern, e.g., a variable name, and the \emph{signified} is another bit pattern that is designated, indirectly or directly, by the signifier.}

We distinguish between three different signifiers in the CS: 1) Bare variable name, 2) Literal, or 3) Meaningful variable name.

\subsubsection{Bare variable name as Signifier}
\label{sec:org71663ee}
The sole function of a bare variable name is to identify a referent in the CS, i.e., a memory location in the computer, that houses another bit pattern called the \emph{value} denoted by the signifier. A bare variable name has no meaning outside the CS.  Values, i.e., the signified bit-patterns, considered in isolation \emph{without an explanation} that informs about the context (e.g., entity, property, time, and space), are sufficient to provide an input to a computation but do not make sense to a human. A sequence of timeless mathematical and/or logical operations of a computer that manipulate input bit-patterns (values) and produce result bit-patterns is called an \emph{algorithm}.  Since a particular mathematical or logical operation requires an appropriate structure of the accessed bit-patterns, a \emph{type} is assigned to the bit-pattern that is used in a computation. The type informs about the structure of the bit-pattern and makes it possible for a compiler to establish the compatibility between the bit-pattern and the operations performed with this bit-pattern.

\subsubsection{Literal as Signifier}
\label{sec:orga975dca}
A literal is a bit-pattern that denotes a specific value (a constant) that is derived from the gestalt of the bit-pattern in the given context.

\subsubsection{Meaningful variable name as Signifier}
\label{sec:org9d4ffc9}
A variable name that is used in the CS and has a meaning in the PS establishes a link between the bit-pattern (value) in the CS and an entity in the PS. It thus provides an \emph{explanation} of the value and brings the temporal aspects of the PS into play. In a database, the \emph{database schema} provides the links between the values stored in the database (in the CS) and the real-world entities and their properties in the PS \cite[pp.~30-35]{kopetz2022data}.

\medskip
\textbf{Example:} Let us assume that a CS variable with the meaningful name \emph{speed} has a value of 25. This variable implies the proposition

\begin{center} speed \emph{is} 25. \end{center}

at the instant of uttering the ITOM (or accessing a speed sensor).  In the PS, the word speed is associated with the dimension [m/sec].  This proposition has thus two temporal aspects, the notion of speed and the instant of utterance (or the instant of sensing the speed). \P
\medskip

Such a meaningful variable name is an instance of an information item (ITOM), which is discussed below.

\subsection{Information Item}
\label{sec:org58ca92d}
\textbf{An \emph{information item} (abbreviated \emph{ITOM}) is the \emph{sense} (the \emph{idea} or \emph{intention}) of a natural language proposition that contains \emph{multiple terms} and temporal parameters that denote the time when the proposition is valid.}

A term can be a \emph{data item} or an \emph{information item (an ITOM)}.  ITOMs are thus bootstrapped, starting with multiple data items. We call an ITOM with data items only a \emph{basic} ITOM. A basic ITOM can be a term in a further ITOM.  A proposition provides information --- an explanation --- about an entity in the PS if at least one of the data items deployed in an ITOM is grounded in the PS.

A typical natural language proposition consists of three terms, a \emph{subject}, a \emph{predicate}, and an \emph{object} that is uttered at a particular instant.

\begin{center}
<subject term> <predicate term> <object term>
\end{center}

As explained before, a term can be a \emph{data item} or an \emph{ITOM}.  In a natural language proposition, the predicate of a proposition discloses the \emph{timing} of the relation by the use of tense (in German, the predicate is called a \emph{Zeitwort}). Thus, the \emph{predicate term} establishes a time-sensitive link, a timed relation, between the subject term and the object term. There are two temporal aspects to consider in a natural language proposition: the \emph{instant of utterance} and the \emph{tense} of the predicate.  In order to make the temporal aspects of a natural language proposition more precise, we introduce a validity time (v-time) as part of the ITOM.  The validity-time (v-time) is an instant denoted by a single timestamp if the ITOM informs about an event (e.g., an observation or a setpoint) or an interval denoted by two timestamps if the ITOM is valid during an interval. An ITOM has thus the following form

\begin{center}
<v-time> <subject term> <predicate term> <object term>.
\end{center}

In contrast to data items, which are mostly static, information ITOMS are mostly dynamic and require a \emph{v-time} to denote the temporal validity of the ITOM.

An \emph{ITOM} establishes timed links among its multiple terms in the mind of the receiver and thus possibly extends the conceptual landscape of the receiver. If these links are already present in the conceptual landscape of the receiver, then the received \emph{ITOM} is not new to this receiver (but may be new to another receiver). The sense of an \emph{ITOM} can only be understood by a receiver if all terms that are part of the ITOM are familiar to the receiver. 

\subsection{Truth and Falsehood of an ITOM}
\label{sec:orge74a285}
At any given instant (or during a given validity interval), an \emph{ITOM} that informs about a phenomenon in the real world can be TRUE or FALSE.  It is TRUE at this instant \emph{if} the \emph{sense} communicated by the \emph{ITOM} can be verified to \emph{correspond} with reality; otherwise, the \emph{ITOM} is FALSE.  A TRUE \emph{ITOM} that informs about some aspect of the state of the world at a given instant or during a given interval is often called a \emph{fact} and establishes a link between nodes of a \emph{knowledge graph} \cite[pp.32-34]{kopetz2022data}.

Let us summarize the main difference between a data item and an information item (an ITOM) as follows:  A data item is a symbol that consists of a \emph{signifier} and a \emph{signified}. In the PS, the signifier of a data item is a physical pattern (e.g., a sequence of letters or the utterance of a word), and the signified is the time-sensitive meaning of this pattern in the mind of a human.  In the CS, the signifier of a data item is a bit pattern (a name of a memory location in the computer), and the signified is another bit pattern (the value) in the referenced memory location. An information item is the sense of a proposition that can link a time-sensitive data item in the PS with a value in the CS and is valid at a particular instant or during a particular time interval. 

\textbf{Inside the CS, we manipulate data items, but at the interfaces between the CS and the PS, we deal with time-sensitive information items.}

The following example explains this fundamental difference between the \emph{meaning} of a
\emph{data item} and the \emph{sense} of an \emph{ITOM} (information item).

\begin{figure}[htbp]
\centering
\includegraphics[width=9cm]{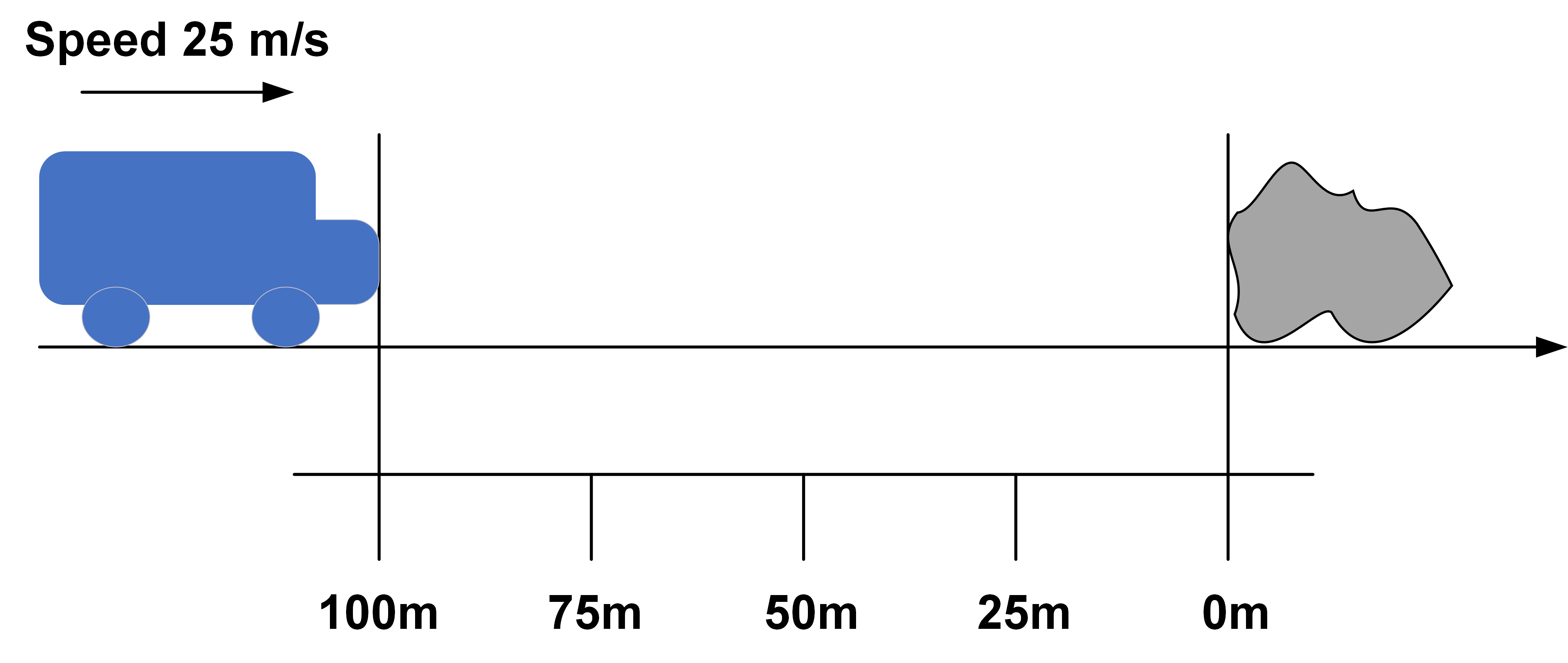}
\caption{\label{fig:org513705d}Car before a rock on the road}
\end{figure}

\medskip
\textbf{Example:} Let us assume that the episode, depicted in  Fig. \ref{fig:org513705d}, starts at 10:59:56 hours. The value 10:59:56 (hours:minutes:seconds) denotes the instant in units of seconds at the beginning of a frame. 
From 10:59:55 to 10:59:57  (an interval), the \emph{blue van from the garage} (a grounded data item) moves with a constant speed of 25 (a literal denoting a value) m/sec (a dimension)  towards a rock on the road (a natural language term). The distance to the rock is measured by a radar system onboard the car. At 10:59:56, the distance of the car from the rock is 100 (a literal denoting a value) m (a dimension). \P
\medskip

This description introduces the following two basic ITOMS, \textbf{under the simplifying assumption that the latencies of the sensors can be neglected}:

\begin{table}[htbp]
\caption{Speed ITOM}
\centering
\small
\begin{tabular}{c|c|c|c}
From 10:59:55 & the blue van & moves to a rock & 25\\
to 10:59:57 & from the garage & with a speed of & [m/sec]\\
\hline
<v-time> & <Subject> & <Predicate> & <Object>\\
\hline
Term\_1 & Term\_2 & Term\_3 & Term\_4\\
\end{tabular}
\end{table}

\begin{table}[htbp]
\caption{Distance ITOM}
\centering
\small
\begin{tabular}{c|c|c|c}
At 10:59:56 & the blue van & is distant & 100\\
 & from the garage & from the rock & [m]\\
\hline
<v-time> & <Subject> & <Predicate> & <Object>\\
\hline
Term\_1 & Term\_2 & Term\_3 & Term\_4\\
\end{tabular}
\end{table}

The ITOMs presented above are basic ITOMs that contain the following data items (Table \ref{tab:orgc7bafb4} and Table \ref{tab:org4e3af37}).

\begin{table}[htbp]
\caption{\label{tab:orgc7bafb4}Data items of the Speed ITOM}
\centering
\small
\begin{tabular}{|l|c|c|c|c|}
\hline
Terms in ITOM & Term\_1 & Term\_2 & Term\_3 & Term\_4\\
\hline
Part of ITOM & v-time & Subject & Predicate & Object\\
\hline
Explanation & Temporal & car & moves & speed\\
 & Validity &  & with & \\
\hline
Dimension & h:min:sec & --- & --- & m/sec\\
\hline
Value & 10:59:55- & --- & --- & 25\\
 & 10:59:57 &  &  & \\
\hline
\end{tabular}
\end{table}

\begin{table}[htbp]
\caption{\label{tab:org4e3af37}Data items of the Distance ITOM}
\centering
\small
\begin{tabular}{|l|c|c|c|c|}
\hline
Terms in ITOM & Term\_1 & Term\_2 & Term\_3 & Term\_4\\
\hline
Part of ITOM & v-time & Subject & Predicate & Object\\
\hline
Explanation & Temporal & car & is & distance\\
 & Validity &  & distant & from the\\
 &  &  &  & rock\\
\hline
Dimension & h:min:sec & --- & --- & m\\
\hline
Value & 10:59:56 & --- & --- & 100\\
\hline
\end{tabular}
\end{table}

When looked at in isolation (without the context provided by the other three data items), a data item has a \emph{meaning} (in the PS) or denotes a \emph{value} (in the CS) but does not tell you anything about the situation of the car.   This information --- the sense of the ITOM --- emerges when the four data items are linked and form an atomic complex, an ITOM, that is valid at the indicated time. 

\section{Temporal Consistency of an Information Item (ITOM)}
\label{sec:org85f1e04}
\label{orgcbb5c25}

Let us assume that the CS observes with its sensors relevant parts of the state of the PS at the beginning a frame, executes a control algorithm (that uses these sensor readings) during the frame and outputs the result, a new setpoint, to an actuator in the PS at the beginning of the next frame. Assume that a frame has a duration of one second. 

The meaning of the bit-pattern acquired by the sensor is determined by the design of the sensor and the physical position of the sensor in the PS. The sensors write these bit-patterns and the observation instants into memory locations of the CS that carry variable names. From the point of view of the CS algorithm, it is sufficient for the variable names in the CS to be \emph{bare (ungrounded) names} that have as referents the cited memory locations with the stored values. \textbf{The grounding of a variable name is not required for the execution of an algorithm in the CS.} It is, however, helpful to a human engineer who wants to understand the functions of an algorithm. Understanding advances if a variable name of the CS is grounded in the PS and the bit-pattern acquired by a sensor is transformed to a value in a standardized dimension.

Let us now assume that the CS performs the following calculations of the brake force and the time it takes the car to stop before the rock, using the \emph{speed ITOM} and the \emph{distance ITOM} introduced in Section \ref{orgaf2545e} under the simplifying assumption, that the latency of the sensors and actuators can be neglected:

\begin{table}[htbp]
\caption{CS Calculations}
\centering
\footnotesize
\begin{tabular}{|c|c|c|c|c|}
\hline
At & Itom & Formula & Dimension & Value\\
 &  &  & Check & \\
\hline
10:59:56:00 & Brake & \(v^2/2s=-d\) & \([m^2/sec^2]/[m]\) & \(-625/200\)\\
 & force &  & \(=[m/sec^2]\) & \(= -3.12\)\\
 &  &  &  & \([m/sec^2]\)\\
\hline
10:59:56:00 & Time & \(v/d=t\) & \([m/sec]/[m/sec^2]\) & \(25/3.12\)\\
 & to stop &  & \(= [sec]\) & \(=8 [sec]\)\\
\hline
\end{tabular}
\end{table}

The \emph{result ITOM} in Table \ref{tab:orgadfbe18} informs about the results of these calculations.

\begin{table}[htbp]
\caption{\label{tab:orgadfbe18}Result ITOM}
\centering
\small
\begin{tabular}{c|c|c|c}
At 10:59:56 & the brake force that & is & \(-3.125\)\\
 & must be applied &  & \([m/sec^2]\)\\
 & to stop the car &  & \\
 & before the rock &  & \\
\hline
<v-time> & <Subject> & <Predicate> & <Object>\\
\hline
Term\_1 & Term\_2 & Term\_3 & Term\_4\\
\end{tabular}
\end{table}

\begin{figure}[htbp]
\centering
\includegraphics[width=8cm]{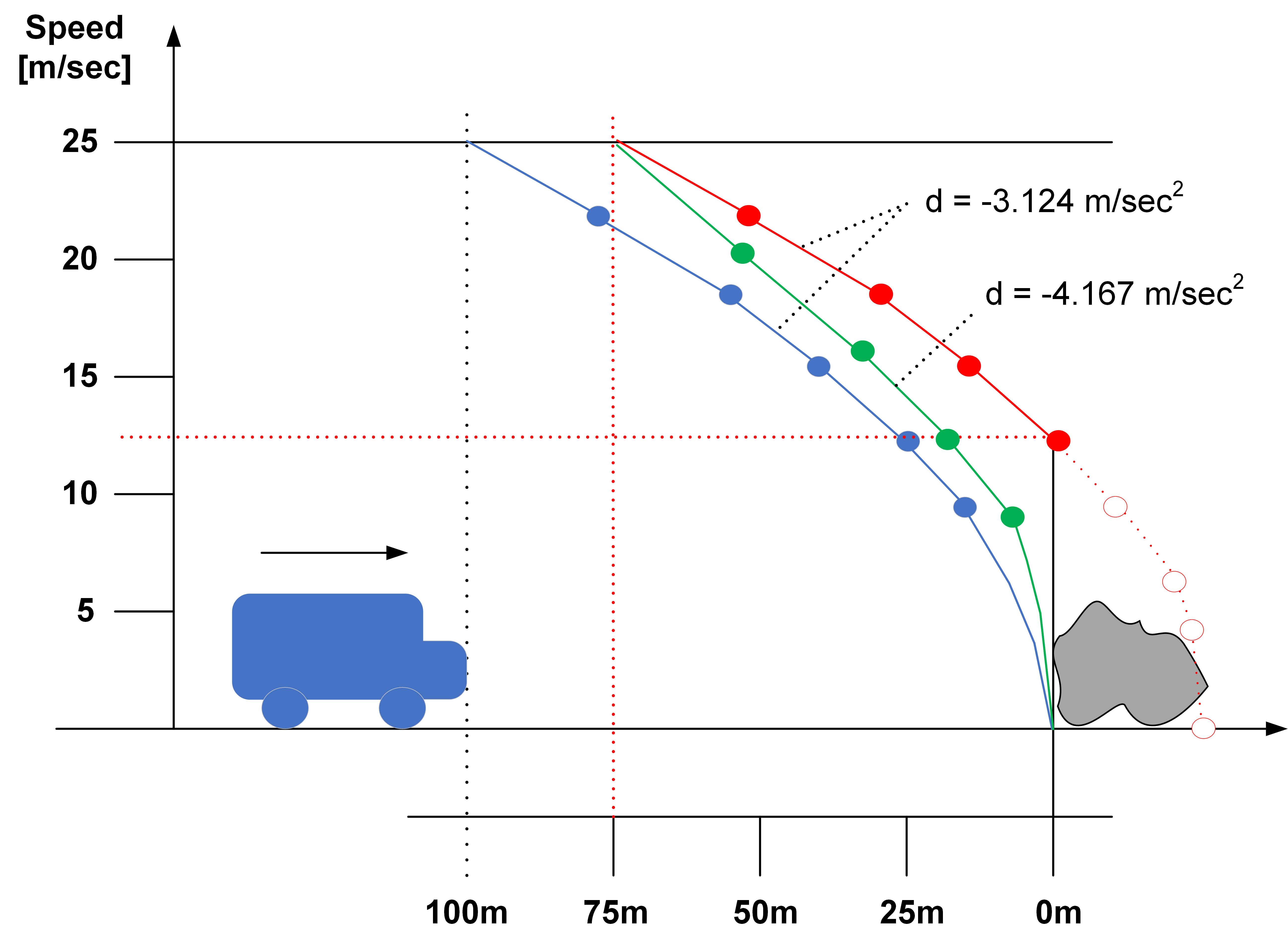}
\caption{\label{fig:org6f4a73c}Speed versus distance of a car as a function of a braking action.}
\end{figure}

This \emph{result ITOM} is temporally consistent, given the calculated set-point, the \emph{brake force}, was applied in the PS at the beginning of the frame, the instant 10:59:56 (blue line in Fig. \ref{fig:org6f4a73c}). However, the calculated setpoint is not available in the PS before the beginning of the next frame (note that a frame has a duration of one second). At the beginning of frame 10:59:57, the \emph{distance ITOM} that informs about the position of the car at 10:59:56 is no longer valid. This makes the \emph{result ITOM} temporally inconsistent because the brake is not applied at the beginning of 10:59:56.  In case we apply the brake force calculated for the instant 10:59:56 at the later instant 10:59:57, the car will hit the rock with a speed of about 45 km/h (red line in Fig. \ref{fig:org6f4a73c}).

This brings us to the following definition of the \emph{temporal consistency} of an ITOM:

\textbf{An ITOM is \emph{temporally consistent} if all terms that are referenced (directly or indirectly) by this ITOM are \emph{TRUE} at the v-time of the ITOM.}

The v-time of a setpoint ITOM is the \emph{interaction instant} when the output ITOM, i.e., the setpoint calculated in the CS, becomes effective in the PS.

\section{Solution Method}
\label{sec:org8bad37c}
\label{orgd66a559}

The problem of the inconsistency of an output ITOM arises because there is a non-negligible interval (we call it the \emph{action interval}) between the \emph{input interaction instant} --- the \emph{instant} when a sensor value is TRUE in the environment (the PS) --- and the \emph{output interaction instant} --- the \emph{instant} when the actuator value is realized in the environment (the PS).  The action interval can be reduced by increasing the performance of the computer in the CS, but it can never be fully eliminated.

One way to solve this problem of inconsistency is the extrapolation of a measured sensor value from the \emph{input interaction instant} to an estimated future value at the \emph{output interaction instant}, \textbf{before} the computation of the result (setpoint) starts in the CS. These estimated values are then used to calculate the setpoint.  This solution method requires that the future interaction instants are known \emph{a priori}. An \emph{anytime algorithm} \cite{kopetz2018anytime} can be used in the computation in case the WCET (worst-case execution time) of an exact computation is unpredictably long. An anytime algorithm is an algorithm that consists of a root segment and an iteration segment that improves the result produced by the root segment. The iteration segment is repeatedly executed, improving the result until an apriori known deadline. 
In an event-triggered system, the output interaction instant depends on the input-data-dependent duration of the computation of the setpoint value.   It is not possible to extrapolate the input data precisely before the computation starts in a purely event-triggered system because at the instant when the computation starts, the future output interaction instant is not yet known in an event-triggered system. Thus, the extrapolation error is higher in an event-triggered system than in a time-triggered system. It depends on the concrete CPS if this higher extrapolation error can be tolerated, or not. 

One well-known technique for this extrapolation is the Taylor expansion.  In a time-triggered CPS the input data from the PS are acquired at the beginning of every new frame. It follows that a time-series of input data of the recent past is available when a new frame starts. This time series can be used to estimate the derivates of the different orders that are required in a Taylor expansion.

\begin{equation}
f(t_1) = f(t_0) + \frac{f'(t_0)}{1!} (t_1-t_0) + \frac{f''(t_0)}{2!} (t_1-t_0)^2 + ...
\end{equation}

Let us now look, in order to get the point across, to our example, where the car moves with a constant speed of 25 m/sec towards the rock. In this example the Taylor expansion stops at the first order derivative. The expansion leads to a distance of the car to the rock of 75 m at the beginning of the frame 10:59:57.         

The following new distance ITOM contains this adjusted distance of 75 m at 10:59:57

\begin{table}[htbp]
\caption{New Distance ITOM}
\centering
\small
\begin{tabular}{c|c|c|c}
At 10:59:57 & the car with a & is distant & 75 m.\\
 & speed of 25m/sec & from the rock & \\
\hline
<v-time> & <Subject> & <Predicate> & <Object>\\
\hline
Term\_1 & Term\_2 & Term\_3 & Term\_4\\
\end{tabular}
\end{table}

The \emph{speed ITOM} of Section \ref{orgaf2545e} is valid during the interval 10:59:55 to 10:59:57, i.e., it is still valid at the instant 10:59:57. The extrapolated distance ITOM for the instant at the start of 10:59:57 is used to calculate (during the frame starting at 10:59:56) the brake force that is applied to the brakes in the PS at 10:59:57. The new result ITOM is temporally consistent because all constituent terms are valid at 10:59:57.

\begin{table}[htbp]
\caption{New CS Calculations}
\centering
\footnotesize
\begin{tabular}{|c|c|c|c|c|}
\hline
At & Itom & Formula & Dimension & Value\\
 &  &  & Check & \\
\hline
10:59:57 & Brake & \(v^2/2s=-d\) & \([m^2/sec^2]/[m]\) & \(-625/150\)\\
 & force &  & \(=[m/sec^2]\) & \(= -4.167\)\\
 &  &  &  & \([m/sec^2]\)\\
\hline
10:59:57 & Time & \(v/d=t\) & \([m/sec]/[m/sec^2]\) & \(25/4.167\)\\
 & to stop &  & \(= [sec]\) & \(= 6 [sec]\)\\
\hline
\end{tabular}
\end{table}

\begin{table}[htbp]
\caption{New Result ITOM}
\centering
\small
\begin{tabular}{c|c|c|c}
At 10:59:57 & the brake force that & is & \(-4.167\)\\
 & must be applied &  & \([m/sec^2]\)\\
 & to stop the car &  & \\
 & before the rock &  & \\
\hline
<v-time> & <Subject> & <Predicate> & <Object>\\
\hline
Term\_1 & Term\_2 & Term\_3 & Term\_4\\
\end{tabular}
\end{table}

A brake force of 4.167 \([m/sec^2]\)  applied at the start of frame 10:59:57  brings the car to a stop before the rock (green line in Fig. \ref{fig:org6f4a73c}). The calculations have been performed under the simplifying assumption that the latencies of the sensors and the actuators can be neglected.

\begin{figure}[htbp]
\centering
\includegraphics[width=9cm]{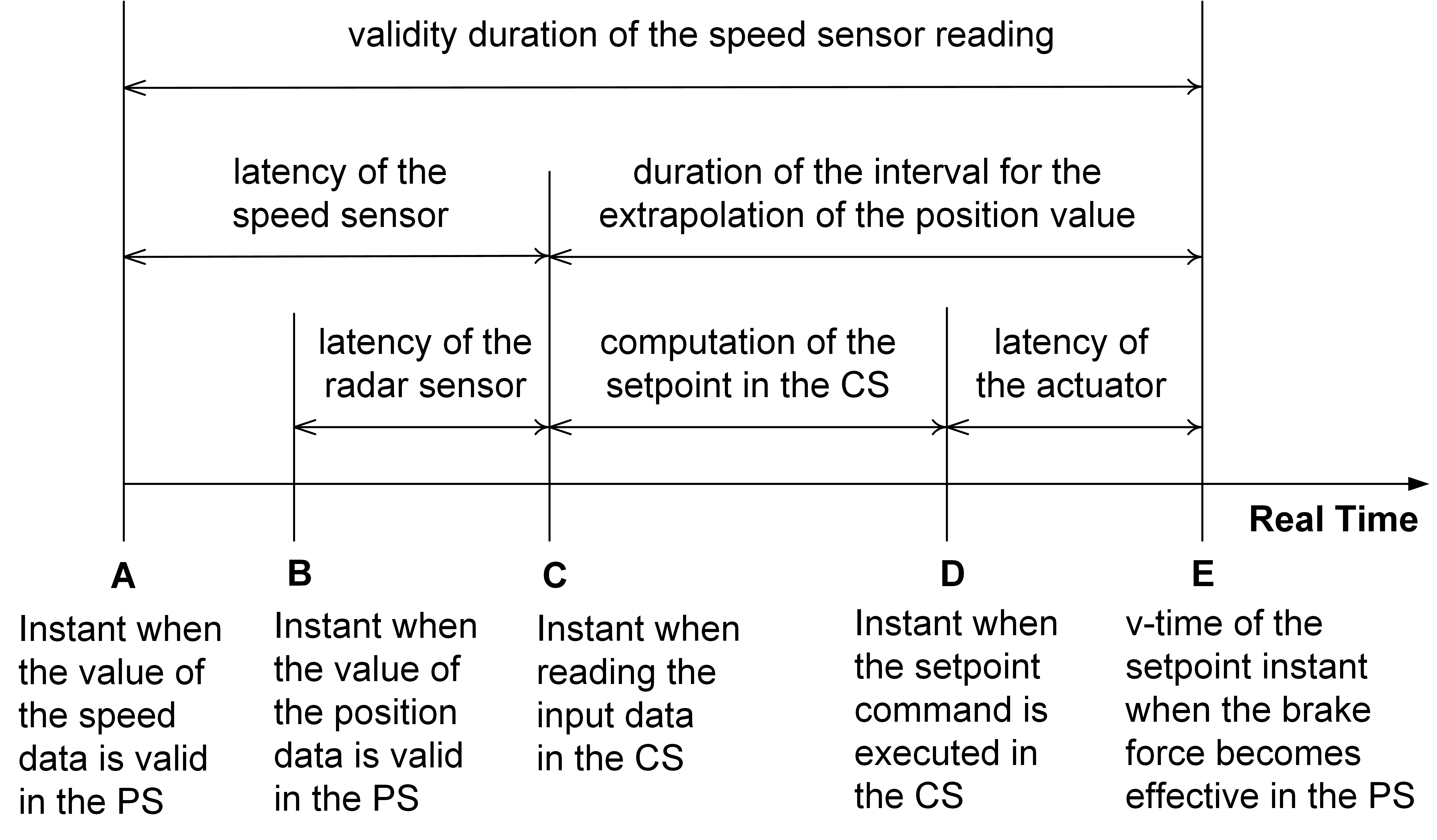}
\caption{\label{fig:org0e0e534}Timing with sensor and actuator latencies}
\end{figure}

In many CPS applications this simplifying assumption of zero latency of the sensors and actuators cannot me made, because the latency of a sensor at the interface between the PS and the CS or the latency of the actuator at the interface between the CS and the PS is significant in relation to the dynamics of the physical process. Fig. \ref{fig:org0e0e534} depicts the full picture, where all latencies of the sensors and actuators are considered.

The notion of sensor \emph{latency} refers to the duration between the instants when an acquired input value is TRUE in the PS (instants A and B in Fig. \ref{fig:org0e0e534}) and the instant when these values are offered for reading in the CS (instant C in Fig. \ref{fig:org0e0e534}). In a time-triggered system, a sensor value is offered to the software in the CS at the SOF.  It is good practice that the software module that deals with a sensor transforms the bit-pattern acquired from the sensor to standardized engineering units also performs an extrapolation from the instants when the values of the sensors are TRUE in the PS (instants A and B in Fig. \ref{fig:org0e0e534})  to the value that is TRUE at the SOF (instant C in Fig. \ref{fig:org0e0e534}).

The notion of \emph{actuator latency} refers to the duration between the instants when the output command with the setpoint value is executed in the CS and the instant when this output command becomes active in the PS (interval between instants D and E in Fig. \ref{fig:org0e0e534}). Given that this actuator latency is known \emph{a priori}, the extrapolation of the sensor values to the v-time of the output command (instant E in Fig. \ref{fig:org0e0e534}) can be performed in the software module that calculates the setpoint value.

In summary, the \emph{action interval}, i.e., the interval between the \emph{input interaction instant} and the \emph{output interaction instant}, consists of the sum of the following three intervals:
\begin{enumerate}
\item The \emph{latency of the sensor}, i.e. the interval between the instant when the input information item is TRUE in the PS and the instant when the respective input value is accessed in the CS.
\item The \emph{duration of the computation}, i.e., the interval between the instant when an input value is accessed in the CS and the instant when the respective output command is executed in the CS.
\item The latency of the actuator, i.e., the interval between an output command executed in the CS and the instant when the respective information item becomes TRUE in the PS.
\end{enumerate}

\section{Temporal Consistency in a Co-Simulation of a CPS}
\label{sec:org719679a}
\label{orgfe587a8}

The global behavior of a CPS emerges from the interactions of two loosely coupled subsystems, the PS and the CS. The interactions between these subsystems are realized by the transfer of ITOMS at recurring interaction instants. The information flow from the PS to the CS occurs at the \emph{input interaction instants}. The information flow from the CS to the PS occurs at the \emph{output interaction instants}. We can model the global behavior of a CPS by the composition of two independent black-box simulation models, one model for the PS and the other model for the CS, that interact at the above-mentioned interaction instants.

If we assume that the temporal behavior of the PS-simulation model at the interaction instants is identical to the behavior of the PS we arrive at a \emph{real-time simulation model} of the PS.  In a real-time simulation model the rate of progress of the model-time corresponds to the rate of progress of physical time.

In a number of CPS applications, it is advantageous if models are developed where the model-time rate can be increased or decreased in relation to physical time. If the model time rates are increased w.r.t. physical time, then it is possible to study the future behavior of the CPS in a shorter physical time span.

In a co-simulation of a CPS, the chosen model times have to proceed at the same rate in the PS model and the CS model.  The relevant input ITOMs acquired from the PS model have to be altered before the setpoint computation is started in the CS model in order to comply with the temporal consistency requirement of the future setpoint ITOM delivered from the CS model to the PS model. Let us assume that an input ITOM is a proper representation of the state of the PS model at the given input interaction instant (in the chosen model time). This input ITOM has to be extrapolated to the instant or the future setpoint ITOM, the output interaction instant (in the chosen model time) before the relevant computation is started in the CS model in order to arrive at a temporally consistent setpoint ITOM.

\section{Consequences for the CS Design}
\label{sec:orgd67ba59}
\label{org5372be1}

The following six design principles help to handle the problem of the temporal inconsistency of the output ITOM.

\textbf{Global Time}: In a dependable distributed CS a fault-tolerant global time with an adequate precision is needed in all subsystems of the CS. This global time provides the capability to interpret a timestamp, produced in one subsystem, in another subsystem of the distributed system.  A precision is adequate if the granularity of the global time is sufficient to record the fine dynamics of the PS process. 

\textbf{Determination of the precise interaction instants}:  On the input side, the \emph{input interaction instant} is the point in time when a sensor value acquired by the CS represents the physical quantity in the PS. On the output side, the \emph{output interaction instant} is the point in time when the actuator value supplied by the CS to the PS is realized in the PS. A detailed analysis of the physical design and functioning of the sensor or actuator is necessary to establish the relevant latencies of sensors and actuators. The difference between the future output interaction instant of the setpoint and the input interaction instant of the input data is the parameter that is required for the extrapolation of the input data, e.g., by a Taylor expansion.

\textbf{Time-triggered operation}:  The sensor values acquired at input interaction instants must be adjusted by extrapolation to the future output interaction instant before the computation of the setpoint starts. This is straightforward in a time-triggered system, where the periodically occurring interaction instants are known in advance.

\textbf{Recording the time series of the sensed values}: In a time-triggered system, the sensor readings are performed periodically at regular instants. The acquired values of the PS should be stored in the CS so that an estimate of the first order derivative, and possibly the second order derivative of a sensor value can be calculated in the CS.

\textbf{Anytime Algorithms}: Properly designed anytime algorithm should be used in the CS to have a satisfying (not necessarily optimal) result available at the \emph{a priori} known interaction instant.

\textbf{Grounding of the Variable Names}:  In order to improve the understandability of the CS software, the following aspects should be considered when naming a variable in a CS program. A name of a variable has:
\begin{enumerate}
\item to denote a memory location in the CS that contains the referenced value of the named variable,
\item to specify how the addressed data item must be interpreted by an algorithm. This is achieved by assigning a \emph{type} to the variable name, and
\item to explain the meaning of the bit pattern by grounding the variable name in the context of the PS.
\end{enumerate}
Furthermore, the acquired bit pattern delivered by a sensor should remain hidden within the sensing module that transform the bit pattern to values of standard engineering units.

\section{Related Work}
\label{sec:orgeb56e74}
\label{org0c165d0}

The importance of time in computer science has long been realized; Furia et al. \cite{furia2012modeling} give a broad overview of timing models. In this paper, we do not introduce a new time model but argue that any CPS is best described by two different time models, one for the PS and another one for the CS. These two models of time must be aligned. Otherwise, temporal inconsistencies may occur, as we have exemplified in this paper.  

The particular problem of sensor input latency in CPS is addressed, for example, in Brahmi et al. \cite{brahmi2013} or Mehmed et al. \cite{mehmed2017deterministic}. However, handling sensor latency is only one aspect of ensuring temporal consistency, as discussed in this paper. For example, while the related work addresses how to calculate (and minimize) the inaccuracy of the sensor data when the sensor data first enters the CS of the CPS, we propose a method to extrapolate the future inaccuracy of the sensor data at the \emph{interaction-instant} when said input (or functions thereof) affects the PS via actuators.

Timing effects are also studied in control theory. For example, Srinivasagupta et al. \cite{srinivasagupta2004time} present a time-stamped model predictive control (TSMPC) algorithm. TSMPC \emph{is able to handle the effect of a control input applied arbitrarily in time.} In our work, we do not require a special control algorithm that mitigates timing-imposed data inaccuracies. In a time-triggered system, we can design the \emph{interaction-instants} between the PS and the CS and, therefore, extrapolate the sensor data for the interaction-instant when the result of the CS calculations is applied to the PS.

\section{Conclusion}
\label{sec:orge6024d6}
\label{org3e261b7}

The problem of the temporal consistency of an output ITOM is acute in CPS applications, where the \emph{action interval} is significant in relation to the dynamics of the physical process in the PS.   This is the case when the process in the PS is changing rapidly, the software in the CS is large and partitioned into a number of communicating subsystems (communication takes time), when the WCET (worst-case execution time) required to compute a result by a complex algorithm, e.g., object recognition by using an AI algorithm, is long and where the sensor latency and the actuator latency are significant.

One solution of this problem relies on the extrapolation of the acquired input data from the instant of data acquisition to the instant when the computed setpoint is used in the environment, the PS. This requires that the instant when a setpoint will be used in the PS is known in advance.

\bibliography{biblio-temp-consistency}
\end{document}